\documentclass[twocolumn,showpacs,preprintnumbers,amsmath,amssymb]{revtex4}
\usepackage{graphicx}
\usepackage{dcolumn}
\usepackage{bm}

\newcommand{\fif}{$^{15}$ND$_3$}
\newcommand{\JK}{$\left|J,K\right>=\left|1,1\right>$}

\begin{document}
\title{An AC electric trap for ground-state molecules}
\author{Jacqueline van Veldhoven$^{1,2}$}
\author{Hendrick L. Bethlem$^{1,2}$}
\author{Gerard Meijer$^{1}$}
\affiliation{
~$^{1}$Fritz-Haber-Institut der Max-Planck-Gesellschaft\\
Faradayweg 4-6, D-14195 Berlin, Germany\\
~$^{2}$FOM-Institute for Plasmaphysics Rijnhuizen, P.O. Box 1207,\\
NL-3430 BE Nieuwegein, The Netherlands}

\date{\today}

\begin{abstract}
We here report on the realization of an electrodynamic trap, capable of
trapping neutral atoms and molecules in both low-field and high-field seeking
states. Confinement in three dimensions is achieved by switching between two
electric field configurations that have a saddle-point at the center of the
trap, i.e., by alternating a focusing and a defocusing force in each direction.
AC trapping of \fif{} molecules is experimentally demonstrated, and the
stability of the trap is studied as a function of the switching frequency. A
1~mK sample of \fif{} molecules in the high-field seeking component of the
\JK{} level, the ground-state of para-ammonia, is trapped in a volume of about
1~mm$^3$.
\end{abstract}

 \maketitle

During the last years, trapping of molecules using optical fields
\cite{Takekoshi:PRL81:5105}, magnetic fields \cite{Weinstein:Nat395:148}, and
electric fields \cite{Bethlem:Nat406:491} has been experimentally demonstrated.
In the largest and deepest traps thus far realized, made of static
inhomogeneous magnetic and/or electric fields, paramagnetic or polar molecules
in low-field seeking quantum states can be confined
\cite{Wing:ProgQuantElectr8:181}. There is a special interest in polar
molecules, as they are promising for a variety of fundamental physics studies
and applications \cite{Bethlem:IntRevPhysChem22:73,TotalIssue:EPJD31}. For most
of these investigations, however, trapping of molecules in high-field seeking
states is required. For instance, only when polar molecules are trapped in
their absolute ground-state, which is high-field seeking for any molecule,
increasing the phase-space density via evaporative cooling is expected to be
possible; the strong dipole-dipole interaction between polar molecules is
predicted to lead to high trap loss rates due to inelastic collisions for
molecules in low-field seeking states \cite{Bohn:pra63:052714}, thereby
hampering the evaporative cooling process. Furthermore, when biomolecules or
heavy diatomic molecules, e.g., those that are important for measurements on
the electric dipole moment of the electron
\cite{Hudson:PRL89:023003,Kawall:PRL92:133007}, are to be trapped, a trap for
molecules in high-field seeking states is the only viable option. Due to their
small rotational energy level spacings, these molecules are purely high-field
seeking already in relatively small electric or magnetic fields. In principle,
optical trapping can be used for molecules in high-field seeking states
\cite{Takekoshi:PRL81:5105,Grimm:AdvAtomMolOptPhys42:95}. In this Letter we
experimentally demonstrate an AC electric trap, a new trap for molecules in
high-field seeking states with a depth and a volume that is considerably larger
than obtainable with an optical trap.

Creating a maximum of the static electric field in free space in three
dimensions, which would allow trapping of molecules in high-field seeking
states, is fundamentally impossible \cite{Earnshaw:TransCambPhilSoc7:97}. In
two dimensions such a field maximum can be created. In a cylindrically
symmetric geometry, for instance, a static electric field can be made with a
maximum in the radial direction and a minimum in the axial direction, or vice
versa. Switching between these two saddle-point configurations results in a
field that is alternatingly exerting a focusing and a defocusing force in each
direction for molecules in high-field (and low-field) seeking states.
Confinement in three dimensions can thus be obtained.

\begin{figure}
\includegraphics[width=\linewidth]{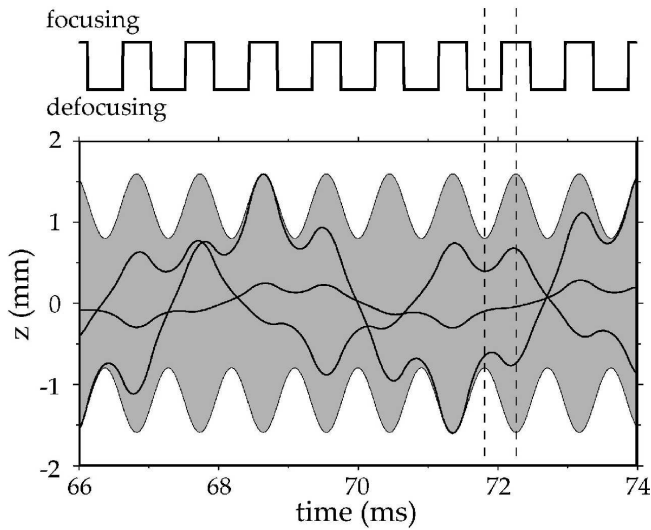}
\caption{Trajectories of molecules along the z-direction of the AC trap as a
function of trapping time, for a switching frequency of 1100~Hz. The block-wave
above the graph indicates whether the electric field configuration is focusing
or defocusing along $\hat{\textrm{z}}$. The grey area shows the size of a
package of molecules, which is seen to be largest when the electric field
configuration is focusing. The two dotted lines are at 79 and 79.5 periods of
switching, when the measurements on the size of the trapped sample shown in
Fig.~\ref{temperature} are performed.} \label{trajectories}
\end{figure}

In our experiment we create the two required saddle-point electric field
configurations either by adding a hexapole field to, or by subtracting a
hexapole field from, a dipole field, as suggested by Peik
\cite{Peik:EPJD6:179}. As switching between the two configurations involves
changing the hexapole field only, the electric field in the center of the trap
doesn't change, which has, for instance, the advantage that Majorana
transitions are prevented. The magnitude of the force on a molecule, which is
proportional to the distance from the center of the trap, is the same in the
two electric field configurations, but the direction of the force is opposite.
At any fixed position, the force on a molecule will average out over time.
However, a molecule does not stay at a fixed position but will move towards the
center of the trap under the influence of a focusing force. It will then be
closer to the center, where the force is smaller, when the defocusing force is
applied. This defocusing force will move the molecule further away from the
center again, bringing it in a region of a larger force when the focusing force
is applied. On average, molecules will thus be further away from the center of
the trap when the focusing force is applied than they are while being
defocused, resulting in a net focusing force. This is shown in
Fig.~\ref{trajectories}, where three trajectories of molecules along the
z-direction are displayed during several switching periods. The molecules
perform a fast micromotion at the switching frequency superimposed on a slower
'secular' motion. The grey area encompasses all stable trajectories, and shows
the size of the package of molecules along the z-direction in the trap as a
function of time. This size is indeed seen to be largest (smallest) when
focusing (defocusing).

Whether the trap is stable or not depends on the (radial) frequency $\Omega$
with which the two electric field configurations are switched. This is most
conveniently discussed when it is assumed that the electric hexapole field is
modulated with a sinusoidal function, rather than with a square wave. The
qualitative behavior of the trap as a function of $\Omega$ will be the same,
but for a sinusoidally varying field the trajectories of the molecules are
solutions of the well-known Mathieu equation
\cite{Peik:EPJD6:179,Abramowitz:HandbookMathFunc}:
\begin{equation}
\label{mathieu} \frac{d^2z}{d \tau^2}+(a-2q_z \cos 2\tau)z=0,
\end{equation}
with
\begin{equation}
\label{cut-off} a= 0;\qquad q_z =
\frac{12\mu_{eff}U_3}{m\Omega^2z_0^3};\qquad\tau = \frac{\Omega
t}{2},
\end{equation}
where $\mu_{eff}$ is the effective dipole moment of the molecule, $U_3$ is the
amplitude of the oscillating hexapole term, $m$ is the mass of the molecule,
and $z_0$ is the characteristic radius of the trap.  When $a=0$, solutions of
the Mathieu equation are stable along the z-direction when $|q_z|<0.907$. If
the voltages on the trap are kept constant, the value of $|q_z|$ can be changed
by changing $\Omega$. At low frequencies, solutions to the Mathieu equation are
unstable and the amplitude of the molecular motion will increase exponentially
in time. Above some cut-off frequency, the trap will abruptly become stable.
When the frequency is increased further, the molecules have less time to move
in between switching times. Their amplitudes during focusing and defocusing
approach each other and the net force on the molecules averages out more. The
trap will thus be deepest for frequencies just above the cut-off frequency and
will become less deep for higher frequencies, whereas no trapping at all is
possible for frequencies below the cut-off frequency. The same arguments and
equations apply for the x- and y-direction, with $q_x=q_y=-q_z/2$.

The operation principle of this AC electric trap for neutral polar molecules is
similar to that of a Paul trap for charged particles \cite{Paul:RMP62:531}.
Using magnetic fields, AC trapping has already been demonstrated for cesium
atoms \cite{Cornell:PRL67:2439} and several proposals for an AC electric trap
for atoms have been put forward
\cite{Peik:EPJD6:179,Shimizu:JpnJApplPhys31:L1721,Morinaga:laserphys4:412}. In
two dimensions the same principle has been used to focus neutral molecules in
high-field seeking states in an alternate gradient (AG) decelerator
\cite{Bethlem:PRL88:133003,Tarbutt:PRL92:173002} and in an AC electric guide
\cite{Junglen:PRL92:223001}. It is noted that in an AG focuser constant
voltages can be applied; the molecules will experience an alternating field due
to their forward velocity.

\begin{figure}
\includegraphics[width=\linewidth]{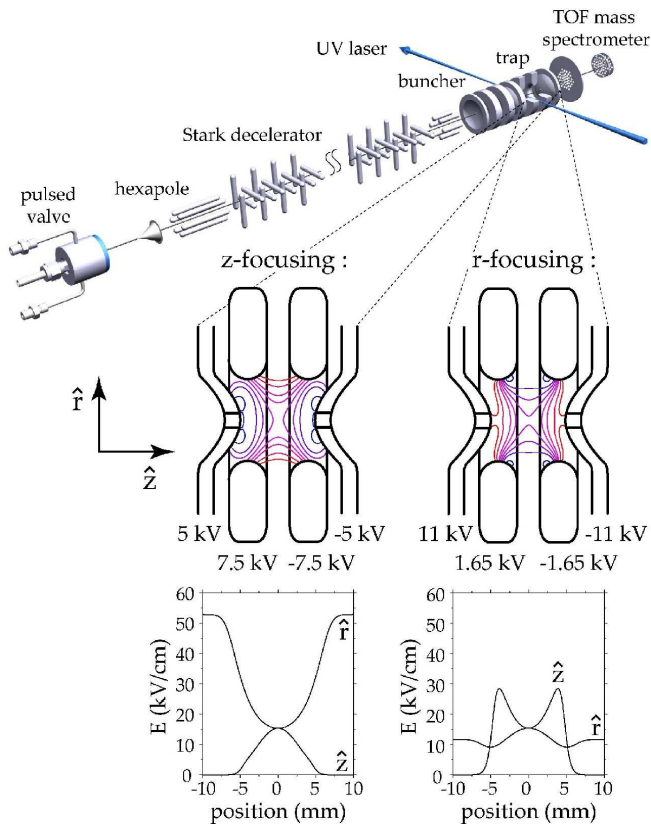}
\caption{Experimental setup. A molecular beam of \fif{} molecules in the
low-field seeking component of the \JK{} level is decelerated and brought to a
standstill in the AC electric trap. Before turning on the trap a transition
from low-field seeking to high-field seeking levels can be made by applying a
microwave pulse. Cross-sections of the trap are shown with lines of equal
electric field strength (contours are shown at 5, 10, 12.5, 15, 17.5, 20, 25
and 30~kV/cm, increasing from blue to red (on-line only)) for both electric
field configurations, together with plots of the electric field strength along
the r- and the z-direction.}\label{setup}
\end{figure}
In the experimental setup shown in Fig.~\ref{setup} a mixture of 5\% \fif{}
seeded in xenon expands from a pulsed valve at a 10~Hz repetition rate. In the
expansion region, about 60\% of the ammonia molecules cool to the \JK{} level,
the ground-state of para-ammonia. After passing through a skimmer, the beam is
coupled into a Stark decelerator by a hexapole. This part of the set-up and the
operation principle of the decelerator have been described in detail elsewhere
\cite{Bethlem:PRA65:053416,Veldhoven:EPJD31:337}. Upon exiting the decelerator
with a forward velocity of around 15~m/s, \fif{} molecules in the low-field
seeking component of the \JK{} level are transversally and longitudinally
focused into the AC trap by a second hexapole and a buncher
\cite{Crompvoets:PRL89:093004}, respectively. A cross-section of the AC
electric trap together with the voltages applied to the electrodes for the two
different electric field configurations is shown in Fig.~\ref{setup} \footnote{
In the experiments reported here, the applied voltages generate a dipole field
with an alternating hexapole field. We have also used this trap with a constant
hexapole field to confine molecules in low-field seeking states.}. The trap has
a hexapole geometry consisting of two ring electrodes with an inner diameter of
10~mm and two end caps, with a closest separation of 9.1~mm. Between the two
ring electrodes there is a 2.9~mm gap through which the UV detection laser is
coupled in. There is a 2~mm diameter hole in the entrance end cap for coupling
the \fif{} molecules into the trap and a 2~mm diameter hole in the exit end cap
for extracting the laser-produced molecular ions. When the package of slow
ammonia molecules enters the trap, voltages are applied to the trap electrodes
such that molecules are brought to a standstill near the center of the trap. At
that time a 20~$\mu$s duration pulse of 1.43~GHz radiation can be coupled in to
induce the transition from the low-field seeking to the high-field seeking
hyperfine levels in the \JK{} state of \fif{} \cite{Veldhoven:EPJD31:337}.
Under optimum conditions, about 20\% of the ammonia molecules are pumped to
high-field seeking levels.

When the AC electric trap is switched on, the voltages on the trap are
alternated at a frequency $\Omega$ between the two configurations shown in
Fig.~\ref{setup}. The configurations on the left-hand side and on the
right-hand side in the figure focus high-field (low-field) seeking molecules in
the axial (radial) and the radial (axial) direction, respectively. The
resulting trapping potential is about twice as deep in the axial direction as
it is in the radial direction. Therefore, increasing the trap depth in the
r-direction at the cost of decreasing the trap depth in the z-direction will
improve the overall trap performance. This can experimentally be done by
adjusting the duty cycle, i.e., the fraction of each switching period during
which the electric field configuration is applied that is denoted as the
z-focusing configuration in Fig.~\ref{setup}. In the measurements the duty
cycle is 45\% when trapping high-field seekers and 55\% when trapping low-field
seekers. The trap is switched on with the r-focusing (z-focusing) configuration
when trapping high-field (low-field) seekers, and the first switching to the
other configuration is performed after a quarter period.

After a certain trapping time, the trap is switched off, and the molecules are
detected using pulsed UV-laser ionization followed by mass-selective detection
of the parent ions. The (2+1)~Resonance Enhanced Multi Photon Ionization
(REMPI) scheme that is used selectively ionizes the \fif{} molecules in the
upper or lower component of the \JK{} inversion doublet, containing the
low-field seeking or high-field seeking levels, respectively.

\begin{figure}
\includegraphics[width=0.75\linewidth]{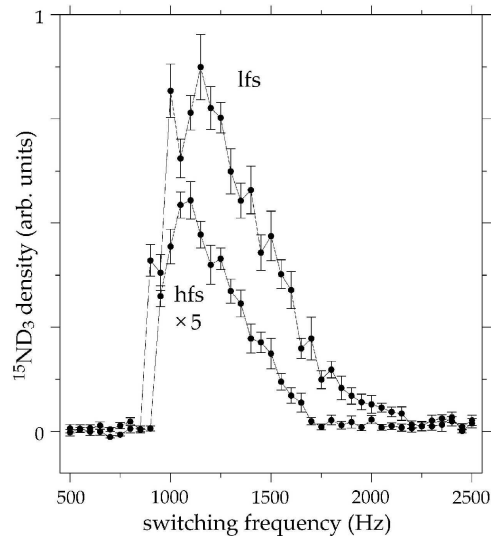}
\caption{Density of \fif{} molecules in low-field seeking (lfs) and high-field
seeking (hfs) levels of the \JK{} state at the center of the trap as a function
of the switching frequency, after the trap has been on for 72~ms. The signal of
the high-field seekers is scaled up by a factor of 5.} \label{frequency}
\end{figure}
In Fig.~\ref{frequency} the density of \fif{} molecules at the center of the
trap is shown as a function of the switching frequency for molecules in
low-field seeking and high-field seeking states. The signal for the high-field
seekers is scaled up by a factor of five, to correct for the 20\% conversion
efficiency in the microwave pumping process. To arrive at the relative
densities, a correction has also been made for the quantum-state specific
ionization efficiency. The measurements are made after the AC trap has been on
for 72~ms, enabling data-acquisition at the 10~Hz repetition rate of the
experiment. The trapping lifetime is actually expected to be several tenths of
seconds, limited by the residual background pressure. The measurements agree
with the qualitative description of the dependence of the stability of the trap
on $\Omega$ as described above. If the cut-off frequency is calculated using
Eq.~\ref{cut-off} with $\mu_{eff}=0.75$~Debye, $z_0\approx4.5$~mm, and
$U_3=3$~kV, a frequency of around 890~Hz is found. The numerically calculated
cut-off frequency for \fif{} in this trap is around 980~Hz, in good agreement
with the measurements. Note that the cut-off frequency for the high-field
seekers is slightly higher than for the low-field seekers due to an about 4\%
difference in $\mu_{eff}$ in the electric field at the center of the trap. With
the present settings, the highest density of trapped molecules is observed at a
switching frequency of 1100~Hz. Due to the anharmonicity of the trap it is
expected that the trap works better for molecules in low-field than in
high-field seeking states, in agreement with the observations.

\begin{figure}
\includegraphics[width=\linewidth]{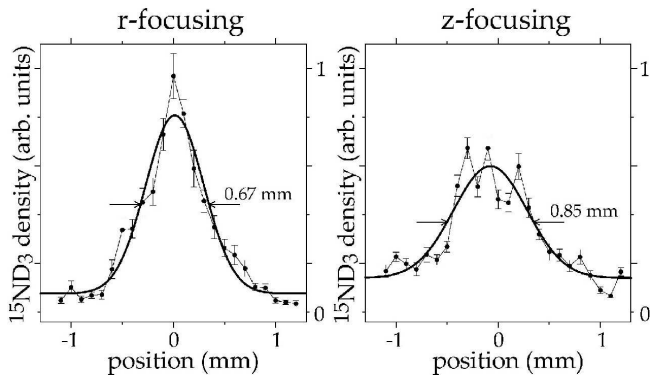}
\caption{Spatial distribution along $\hat{\textrm{z}}$ of \fif{} molecules in
high-field seeking states after trapping at 1100~Hz for 79 (left) and 79.5
(right) switching periods. A Gaussian distribution is fitted to the
measurements and the width (FWHM) of this distribution is given.}
\label{temperature}
\end{figure}
In Fig.~\ref{temperature} measurements of the spatial distribution along the
z-direction are shown for \fif{} molecules in high-field seeking states. The
measurements are performed after trapping for about 72~ms at a switching
frequency of 1100~Hz. The data shown on the left are taken after 79 periods of
switching, i.e., in the middle of r-focusing. The data shown on the right are
taken half a period later, i.e., in the middle of z-focusing. It is clear from
these measurements that the package of molecules in the z-direction is indeed
smaller during r-focusing (denoted as defocusing in Fig.~\ref{trajectories})
than it is during z-focusing (denoted as focusing in Fig.~\ref{trajectories}).
From the observed spatial distributions, a temperature of the trapped sample of
around 1~mK is deduced.

In the AC electric trap demonstrated here, neutral atoms and molecules, both in
low-field and in high-field seeking states, can be trapped. For molecules in
high-field seeking states only optical traps, which have a typical trap depth
below 1~mK and a typical volume of 10$^{-5}$~cm$^3$
\cite{Grimm:AdvAtomMolOptPhys42:95}, had been demonstrated thus far; a
considerably deeper and larger microwave trap has recently been proposed
\cite{DeMille:EPJD31:375}. We have demonstrated here, using \fif{} as a
prototypical polar molecule, an AC electric trap with a trap depth of about
5~mK and a trapping volume of about 20~mm$^3$. This relatively deep and large
trap for ground-state molecules holds great promise for future applications.

\begin{acknowledgments} This work is part of the research program of the
`Stichting voor Fundamenteel Onderzoek der Materie (FOM)', which is financially
supported by the `Nederlandse Organisatie voor Wetenschappelijk Onderzoek
(NWO)'. The technical support by A.J.A. van Roij, the design and construction
of the electronics by G. Heyne and P. Zilske, and discussions with J.
K\"{u}pper and B. Friedrich are gratefully acknowledged.
\end{acknowledgments}


\begin{thebibliography}{10}

\bibitem{Takekoshi:PRL81:5105}
T{.~}Takekoshi, B.M{.~}Patterson and R.J{.~}Knize, Phys. Rev. Lett.
  \textbf{81}, 5105  (1998).

\bibitem{Weinstein:Nat395:148}
J.D{.~}Weinstein, R{.~}{d}e{C}arvalho, T{.~}Guillet, B{.~}Friedrich and
  J.M{.~}Doyle, Nature \textbf{395}, 148  (1998).

\bibitem{Bethlem:Nat406:491}
H.L{.~}Bethlem, G{.~}Berden, F.M.H{.~}Crompvoets, R.T{.~}Jongma, A.J.A{.~}van
  Roij and G{.~}Meijer, Nature \textbf{406}, 491  (2000).

\bibitem{Wing:ProgQuantElectr8:181}
W.H{.~}Wing, Prog. Quant. Electr. \textbf{8}, 181  (1984).

\bibitem{Bethlem:IntRevPhysChem22:73}
H.L{.~}Bethlem and G{.~}Meijer, Int. Rev. Phys. Chem. \textbf{22}, 73  (2003).

\bibitem{TotalIssue:EPJD31}
{Special Issue on \it{Ultracold Polar Molecules}}, Eur. Phys. J. D \textbf{31
  \textnormal{(2)}}  (2004).

\bibitem{Bohn:pra63:052714}
J.L{.~}Bohn, Phys. Rev. A \textbf{63}, 052714  (2001).

\bibitem{Hudson:PRL89:023003}
J.J{.~}Hudson, B.E{.~}Sauer, M.R{.~}Tarbutt and E.A{.~}Hinds, Phys. Rev. Lett.
  \textbf{89}, 023003  (2002).

\bibitem{Kawall:PRL92:133007}
D{.~}Kawall, F{.~}Bay, S{.~}Bickman, Y{.~}Jiang and D{.~}DeMille, Phys. Rev.
  Lett. \textbf{92}, 133007  (2004).

\bibitem{Grimm:AdvAtomMolOptPhys42:95}
R{.~}Grimm, M{.~}Weidem\"uller and Y{.~}Ovchinnikov, Adv. Atom. Mol. Opt. Phys.
  \textbf{42}, 95  (1999).

\bibitem{Earnshaw:TransCambPhilSoc7:97}
S{.~}Earnshaw, Trans. Camb. Phil. Soc. \textbf{7}, 97  (1842).

\bibitem{Peik:EPJD6:179}
E{.~}Peik, Eur. Phys. J. D \textbf{6}, 179  (1999).

\bibitem{Abramowitz:HandbookMathFunc}
M{.~}Abramowitz and I.A{.~}Stegun: \emph{Handbook of Mathematical Functions}.
\newblock 9. edition. Dover Publications, New York, 1970.

\bibitem{Paul:RMP62:531}
W{.~}Paul, Rev. Mod. Phys. \textbf{62}, 531  (1990).

\bibitem{Cornell:PRL67:2439}
E.A{.~}Cornell, C{.~}Monroe and C.E{.~}Wieman, Phys. Rev. Lett. \textbf{67},
  2439  (1991).

\bibitem{Shimizu:JpnJApplPhys31:L1721}
F{.~}Shimizu and M{.~}Morinaga, Jpn.J.Appl.Phys \textbf{31}, L1721  (1992).

\bibitem{Morinaga:laserphys4:412}
M{.~}Morinaga and F{.~}Shimizu, Laser Phys. \textbf{4}, 412  (1994).

\bibitem{Junglen:PRL92:223001}
T{.~}Junglen, T{.~}Rieger, S.A{.~}Rangwala, P.W.H{.~}Pinkse and G{.~}Rempe,
Phys. Rev. Lett.
  \textbf{92}, 223001  (2004).

\bibitem{Bethlem:PRL88:133003}
H.L{.~}Bethlem, A.J.A{.~}van Roij, R.T{.~}Jongma and G{.~}Meijer, Phys. Rev.
  Lett. \textbf{88}, 133003  (2002).

\bibitem{Tarbutt:PRL92:173002}
M.R{.~}Tarbutt, H.L{.~}Bethlem, J.J{.~}Hudson, V.L{.~}Ryabov, V.A{.~}Ryzhov,
  B.E{.~}Sauer, G{.~}Meijer and E.A{.~}Hinds, Phys. Rev. Lett. \textbf{92},
  173002  (2004).

\bibitem{Bethlem:PRA65:053416}
H.L{.~}Bethlem, F.M.H{.~}Crompvoets, R.T{.~}Jongma, S.Y.T{.~}van~de Meerakker
  and G{.~}Meijer, Phys. Rev. A \textbf{65}, 053416  (2002).

\bibitem{Veldhoven:EPJD31:337}
J{.~}van Veldhoven, J{.~}K\"upper, H.L{.~}Bethlem, B{.~}Sartakov, A.J{.~}van
  Roij and G{.~}Meijer, Eur. Phys. J. D \textbf{31}, 337  (2004).

\bibitem{Crompvoets:PRL89:093004}
F.M.H{.~}Crompvoets, R.T{.~}Jongma, H.L{.~}Bethlem, A.J.A{.~}van Roij and
  G{.~}Meijer, Phys. Rev. Lett. \textbf{89}, 093004  (2002).

\bibitem{DeMille:EPJD31:375}
D{.~}DeMille, D.R{.~}Glenn and J{.~}Petricka, Eur. Phys. J. D \textbf{31}, 375
  (2004).



\end{thebibliography}
\end{document}